\documentclass[a4paper,12pt]{pnaslike}
\usepackage[dvipdf]{graphicx}
\usepackage{amssymb,amsfonts,amsmath}

\begin{document}
   
\title{Altruism can proliferate through group/kin selection despite high random gene flow}

\author{
Roberto H. Schonmann\affil{1}{Dept. of Mathematics, University of California at Los Angeles, CA 90095, USA},
Renato Vicente\affil{2}{Dept. of Applied Mathematics, Instituto de Matem\'atica e Estat{\'\i}stica, 
Universidade de S\~ao Paulo, 05508-090, S\~ao Paulo-SP, Brazil}\and
Nestor Caticha\affil{3}{Dep. de F{\'\i}sica Geral, Instituto de F{\'\i}sica, Universidade de S\~ao Paulo, CP 66318, 05315-970, S\~ao Paulo-SP, Brazil}
}

\maketitle

\begin{article}

\begin{abstract} 
The ways in which natural selection can allow the proliferation of
cooperative behavior have long been seen as a central problem in 
evolutionary biology. Most of the literature has focused on interactions 
between pairs of individuals and on linear public goods games.
This emphasis led to the conclusion that even modest levels
of migration would pose a serious problem to the spread of altruism
in group structured populations. Here we challenge this conclusion, by analyzing evolution in a framework which allows for complex  group interactions and random migration among groups. We conclude that contingent forms of strong altruism can spread when rare under realistic group sizes and 
levels of migration. Our analysis combines group-centric and gene-centric perspectives, allows for arbitrary strength of selection, and leads to extensions of Hamilton's rule for the spread of altruistic alleles, applicable under broad conditions. 
\end{abstract}



\keywords{natural selection | population genetics | evolutionary game theory | group selection | kin selection | altruism | Hamilton's rule | iterated public goods game}

\section{1. Introduction}
The evolution of cooperation and altruism are
fundamental scientific challenges  highlighted 
by their role in the major transitions in 
life's history, when natural selection acted simultaneously on several competing levels \cite{MSS, Frank, Rou, LK, Okasha, Nowak, WGF, Bourke}. In this context, the relevance of basic concepts, including group selection and Hamilton's rule 
remain controversial 
\cite{WW, WGG, Foster, Tra, Leigh, NTW1, A100+, GWW, WMG}. Here we address these problems by  studying a framework for evolution in group structured populations
that incorporates inter- and intra-group competition and migration. Combining group-centric with gene-centric perspectives
in a constructive group/kin selection approach, we build methodology 
that allows for the analysis of arbitrary non-linear fitness functions, resulting
from complex multi-individual interactions across life cycles. 
We obtain the conditions for a rare social allele to invade the population. 
This is obtained in a mathematically rigorous way, 
by analyzing the stability of the equilibrium in which this allele is absent. 
This analysis is done for arbitrary strength of selection, but when selection 
is weak and groups are large the condition for invasion simplifies significantly into a form that 
is easy to apply and provides substantial intuition.  
In the case of linear fitness functions, the condition for invasion is 
identical to Hamilton's rule, and it is natural to regard the more general 
non-linear cases as generalizations of that rule. 
Our results also show that one of the most widely used approaches to
analyzing kin selection models, 
\cite{TF}, 
\cite{WGF} (condition (6.7)), and \cite{GWW} (Box 6), yields incorrect results in some biologically relevant situations.

Our results reveal biologically realistic conditions under which 
altruism can evolve when rare, but genetic relatedness in groups is modest.  In this way we challenge a common understanding according to which 
inter-group selection favoring altruism could only override intra-group selection
favoring selfishness
under exceptional conditions, namely small group size and very low migration rates  \cite{Leigh, WMG,MS64, Williams, Ham75, MS76,   
Aoki, CA, Kimura, Leigh83, MS}.  
We identify the 
emphasis on linear public goods games in  
the literature, including most of these papers, as having supported  this belief. In contrast, we show that  for iterated public goods games, in which altruists cooperate 
or not in each round based on previous outcomes 
\cite{Joshi, BR},  altruism can spread even when groups are large, selection is weak and migration rates are substantially larger than the inverse of group size. 
This result corrects \cite{BR}, who predicted that large group size would not allow cooperation to spread when rare in this model. For species that live in groups, several vital group activities repeat themselves
periodically and behavior changes as feedback is obtained from previous iterations. The 
iterated public goods game that we study is therefore often more realistic than a simple 
one shot public goods game. A proper analysis of this model fills therefore an important 
gap in the literature. 
To obtain our result we show that in the absence of selection,
when groups are large, the fraction of group members that are 
close relatives of a randomly chosen individual has a non-Gaussian distribution 
with a fat tail.  As a consequence, even when altruistic alleles are rare in the population,
they have a significant probability of concentrating in some groups, accruing substantial reproductive gains through multi-individual synergy. 

\section{2. The two-level Fisher-Wright framework}

\begin{figure}[h!]
\begin{center}
\includegraphics[width=0.5\textwidth]{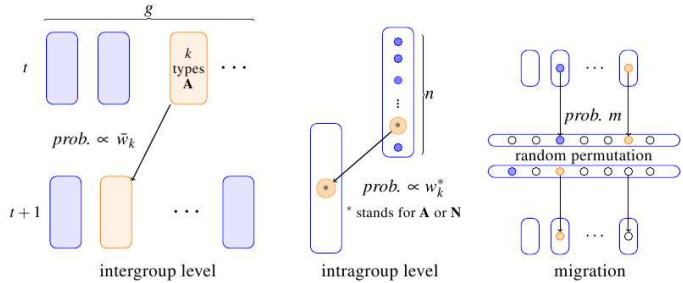}
\end{center}
\caption{  
{\bf Diagram of the 2lFW process.} 
{\bf (Left) FW intergroup competition:}
Each group in the new generation independently descends from
a group in the previous generation,
with probabilities proportional to group average fitness
$\bar w_k = \frac{k w^A_k + (n-k)w^N_k}{n}$. 
{\bf (Center) FW Intragroup competition:} If a group 
descends from  a group with $k$ types A, then it will have $i$ types A with probability $P(k,i) 
= \mbox{bin}(i\, | \, n,k w^A_k / n \bar w_k)$,
where the binomial probability $\mbox{bin}(i|n,q)$ is the 
probability of $i$ successes in $n$ independent trials, each with 
probability $q$ of success. 
{\bf (Right) Migration:}
Once the new $g$ groups have been formed according to the  
two-level  competition process, a random fraction $m$ of the individuals migrates. Migrants are randomly shuffled.  
{\bf Note:}  The assignment of relative fitness to the groups in
the fashion done above is a necessary and sufficient condition \cite{KG-S} for individuals in the 
parental generation to have each an expected number of offspring proportional to their personal relative fitness. 
}
\label{fig1}
\end{figure}

When members of a species live in groups, their reproductive success depends on the behavior of all group members. More efficient groups may grow faster and split, outcompeting 
the less efficient ones that die out.
On the other hand, individuals may free ride on the cooperation of other members of their group,
and in this way outcompete them. 
This picture is further complicated by migration among groups.
The {\it Two-level Fisher-Wright framework with selection and \textsc{}migration} (2lFW) captures all these elements, in a simplified fashion. In 2lFW haploid individuals live in a large number $g$ of groups of size $n$,
and are of two genetically determined phenotypic types, A or N.  
Generations do not overlap,
reproduction is asexual and the type is inherited by the offspring.
(Mutations will be considered briefly later.)  The relative fitness ($w$) of a type A, and that of a type N, 
in a group that has $k$ types A, are, respectively,
$w^A_k = 1 + \delta v^A_k$ and $w^N_k = 1 + \delta v^N_k$,
with the convention that $v^N_0 = 0$, i.e., $w^N_0 = 1$. 
The quantities $v^A_k$ and $v^N_k$ represent life-cycle payoffs  
derived from 
behavior, physiology, 
etc. The parameter $\delta \geq 0$ 
indicates the strength of selection. Fig.~\ref{fig1} describes the creation of a new generation in the 2lFW through  
inter- and intra- group competition, 
followed by migration at rate $m$.

Cases in which types A behave in some altruistic fashion are 
of particular interest \cite{KG-SF}. Most of the literature concerns the very special 
case of a {\it linear} public goods game (PG), defined by $v^A_k = -C + B (k-1)/(n-1)$,
$v^N_k = Bk/(n-1)$, $0 < C < B$,
in which each type A cooperates,
at a cost $C$ to herself, providing a benefit $B$ shared by the 
other members of her group. The need to consider 
more complex intra-group interactions and 
non-linear payoff functions is, nevertheless, well known 
\cite{WGG,NTW1,Joshi, KJC-B, Aviles, BR, HMND,  Vee1, AS, GYO, CRL, SDZ,  
 HP, BJRB, BG, BGB}.
Non-linearities appear naturally whenever activities involve many group members simultaneously. They result from threshold phenomena, increasing returns to scale,
saturation, etc. 
For instance, to hunt large prey may require a large minimum number of hunters, the likelihood
of success may first increase rapidly with the number of hunters, but it may plateau
when this number becomes very large.  
Allowing for the analysis of such synergistic multi-individual interactions and activities  is a central feature of our approach, 
distinguishing it from theoretical frameworks 
based on pairwise interactions, or single actors benefiting a group \cite{LK,Que2, LR}.

The 2lFW framework 
can be seen as a generalization of the trait-group framework
(see Sec. 2.3.2 of \cite{Okasha}), which corresponds to the case $m=1$. One can interpret $1-m$ as an assortment parameter. 
Because migration is completely random in 2lFW, this assortment represents a worst case scenario, abstracting away additional assortment caused by kin 
recognition, greenbeard effects, selective acceptance of migrants,
joint migration of individuals, etc. 
It is well known \cite{Joshi, BR, AS} that even when $m=1$ non-linearities in 
fitness functions allow for coexistence of cooperators and defectors. 
But under the strong altruism condition $v^{A}_{k+1} < v^N_k$
(meaning that each type A would be better off mutating into a type N), this is not 
the case \cite{KG-SF,MJ}. One of our goals is to determine the level of migration 
compatible with invasion by rare strong altruists. 
 
The 2lFW also generalizes the ``typical kin selection model'' of \cite{LKWR}, where the payoffs were those of PG, and the analysis relied on this and on 
the assumption of weak selection. That paper was a response to \cite{TN}, where group selection was argued to be an important mechanism for the evolution
of cooperation, and a multilevel selection model based on Moran's model was introduced. Our analysis of 2lFW with non-linear fitness functions highlights 
the importance of combining group-centric with gene-centric perspectives, and shows that group selection can be an important force in evolution under realistic 
conditions. It also shows that mathematically rigorous analysis can be carried out even when selection is strong and fitness functions are non-linear. And 
it shows that one has to be very careful in applying mathematically non-rigorous methodology, as it can produce substantially incorrect results, even when
selection is weak.

\section{3. A basic example: Iterated public goods game}

\begin{figure}[h!]
\begin{center}
\hspace{-0.5cm}
\includegraphics[width=0.5\textwidth]{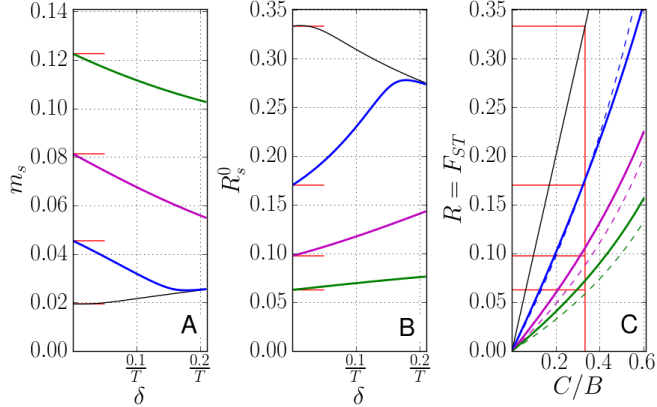}
\end{center}
\caption{
{\bf Iterated public goods game (IPG).} 
A public goods game (PG) is repeated an average of $T$ times in a life-cycle. 
In each round each individual can
cooperate at a cost $C$ to herself, producing a benefit $B$ shared
by the other members of the group.
Types N never 
cooperate, while types A cooperate in the first round and later cooperate only if at least a fraction 
$\alpha$ of group members cooperated in the previous round. 
In all panels
$\alpha = C/B$ (types A are strongly altruistic)
and curves correspond to 
$T = 1$ (black, this case is identical to PG), 
10 (blue), 100 (magenta), 1000 (green) 
(bottom to top in Panel A,
top to bottom in Panels B and C).
{\bf Panel A:} 
$C=1$, $B=3$, $n=50$. 
Curves 
give the
critical migration rate $m_s$ 
below which types A proliferate when rare, and 
that solves $\rho(m_s) = 1$, 
or equivalently $\Delta p = 0$ in (\ref{gen}).
(The subscript `s' stands for `survival'.) The dependence of $m_s$ on 
the strength of selection $\delta$ indicates the relevance of studying both weak and strong 
selection.
Short horizontal red lines indicate value of $m_s$ under weak selection, obtained from setting 
$\Delta p = 0$ in (\ref{ws}) (note the excellent agreement). 
{\bf Panel B:}  Again, $C=1$, $B=3$, $n=50$. Curves give the 
critical relatedness 
$R^0_s = R^0(m_s)$
above which types A proliferate. 
Here $R^0(m) = (1-m)^2/(n - (n-1)(1-m)^2) \approx 1/(1+2nm)$ is the relatedness obtained from neutral genetic markers. 
Short horizontal red lines are again from $\Delta p =0$ in (\ref{ws}). 
{\bf Panel C:} 
Limit of large $n$ under weak selection. Critical values of relatedness $R = F_{ST} = 1/(1+2nm)$,
as function of $C/B$.
Solid lines provide the solution  
to the equation 
$C/B - R = (T-1) R (1-C/B)^{1/R}$ derived from setting $\Delta p =0$ in (\ref{wsln}). Dashed lines
give its approximation (\ref{RCBlnT}). 
Red vertical line corresponds to $C/B = 1/3$,
while horizontal red lines are at the 
same level of those from Panel B. Their intersections illustrate the fact that both the solid and dashed lines 
in Panel C are good approximations to weak-selection values of critical relatedness, $R^0_s$,
when $n = 50$. 
}
\label{fig2}
\end{figure}

Non-linearities in life-cycle payoffs can result from activities repeating themselves during a lifetime, and behavior being contingent on previous outcomes. A basic example is the iterated public goods game (IPG) \cite{Joshi, BR}. In IPG a PG is repeated an average of $T$ times in a life-cycle. We will suppose that types N never cooperate, while types A cooperate in the first round and later cooperate only if at least a fraction 
$\alpha$ of group members cooperated in the previous round.
Mathematically, this model generalizes the iterated prisoner dilemma and tit-for-tat, from the dyadic setting 
of \cite{Trivers} and \cite{AH} to the multi-individual setting. But while direct or indirect reciprocity requires the identification of individuals in the group, this is not the case here. The behavior of types A in the IPG can be triggered by individuals simply discontinuing cooperative behavior when previous cooperation produced negative feedback to them. In other words, allele A can predispose individuals to cooperate, but as they do it
and obtain feedback from that behavior, they may continue it or discontinue it.  The IPG is in this sense closely related to generalized reciprocity mechanisms \cite{RT} with low cognitive requirements. Negative feedback from cooperation  
should occur if the fraction of group members that cooperated was less than $C/B$,  but not if it was larger than that threshold,
since in the former case the payoff to a cooperator is negative, while in the latter case it is positive. This gives a special role to the value $\alpha = C/B$.  If $\alpha \leq C/B$, the behavior of types A is altruistic in the strong sense that each type A individual would increase its fitness if it behaved as a type N, everything else being equal, i.e., $v^A_{k+1} < v^N_k$ (see SI Appendix (Section 8) for a detailed discussion).  Moreover, types N always free ride and have greater fitness than types A in the same group, 
regardless of the values of $\alpha$ and $T$, i.e., $v^A_k < v^N_k$. Fig.~\ref{fig2} displays a detailed analysis of some instances of the IPG, giving conditions for allele A to spread when rare. For many species that live and interact in groups 
for many years, several vital activities, including 
collective hunting and food sharing, can repeat themselves hundreds or thousands of times in a  life-cycle, giving plausibility to the values of $T$ in Fig.~\ref{fig2}.  (The assumption that individuals discontinue behavior after a single unsuccessful participation is a simplification. When this is not a realistic assumption, one can  interpret the parameter $T$ as the ratio between the typical number of repetitions of the activity and the typical number of unsuccessful attempts before 
cooperation is discontinued by a type A.)  Panel C, in which selection is weak and groups are large, shows two important
contrasting results. When $T=1$, and the IPG is identical to the PG, allele A can only invade under Hamilton's condition $R = F_{ST} > C/B$. But as $T$ increases, the level of relatedness needed for invasion drops substantially, so that for modest values of $B/C$, allele A can invade under $R = F_{ST}$
significantly lower than $10\%$, compatible with levels observed in  several species, including humans  \cite{MS} (Table 8.3), \cite{BG}, \cite{HC} (Tables 6.4 and 6.5), \cite{Bowles},  \cite{MHam} (Table 4.9). The corresponding number of migrants per group per generation, $n m = (1-R)/(2R)$, can be of the order of 10. Further examples showing the spread of altruism and cooperation under high levels 
of gene flow and low levels of relatedness are provided in Figs. 5, 6, 7, 8, 20 and 21 in the SI Appendix.

\vspace*{-0.5cm}

\section{4. Condition for invasion under strong selection}
To analyze the 2lFW,
denote by $f_k(t)$, $k =0, ..., n$ the 
fraction of groups in generation $t$ that have exactly $k$ types A.
Denote by $p(t) = \sum_{k=1}^n (k/n) f_k(t)$ the frequency of types A in the population.   
The state of the population in generation $t$ is described by the vector $f(t) = (f_1(t),...,f_n(t))$,
since $f_0(t) = 1 - \sum_{k=1}^n f_k(t)$. We will suppose that $g >> n$, so that, by the law of large numbers, $f(t)$ evolves as a 
deterministic (non-linear) dynamical system in dimension $n$. 
Here we will study its linearization close to the fixed point $(0,...,0)$, with no types A. This means that we are restricting ourselves to the case in 
which $p << 1$, and studying the conditions for allele A to invade the population 
when rare. With notation introduced in Fig.~\ref{fig1}, 
we have then  $f(t+1) = f(t) M (A+B)$, where 
$M_{k,i} = \bar{w_k} \, 
P(k,i)$, $A_{i,j} =
\mbox{bin}(j \, | \, i, 1-m)$ and  $B_{i,j} = m$, if $j=1$, $B_{i,j} = 0$ if $j \not = 1$.
Matrix $M$ represents the production of groups in the new generation, through the two-level competition. Matrix $A$ represents the effect of types A migrating out of groups, and matrix $B$ represents the effect of these migrant types A  joining groups that previous to migration had no types A. (When $p << 1$,  the migrant types A are a small fraction of the migrant 
population, and therefore each one is likely to settle in a different group that had no types A before migration.)  A standard application of the Perron-Frobenius Theorem   
implies that when $t >> 1$, we have, in good approximation,  
$f(t) = C \rho^t \nu$, where $C$ is a constant that depends on $f(0)$, $\rho > 0$ is the leading eigenvalue of 
$M(A+B)$ and $\nu$ is its corresponding left-eigenvector
normalized as a probability vector. This means that, 
regardless of the initial distribution $f(0)$, with 
$0 < p(0) << 1$, demographics and 
natural selection drive $f(t)$ towards multiples of $\nu$, in what can be 
seen as self-organization of copies of A in the optimal stable way for them to spread. Once this has happened, $p(t)$ grows at rate $\rho$. 
Consequently, allele A will  
proliferate, when rare,
if the {\it viability condition}  $\rho > 1$ holds, and it will vanish if $\rho < 1$.   
(See Fig.~\ref{fig2} and   
SI Appendix (Sections 1 and 2)
for applications, illustrations and further explanations.)
When $\rho > 1$, even if allele A is initially absent, a small rate of mutation 
will introduce it, allowing it to then invade the population. 
In the terminology of evolutionary game theory (see, e.g., \cite{MS} Chapter 7), 
phenotype N is an evolutionary stable strategy (ESS) when $\rho < 1$ and N  is not an ESS when $\rho > 1$.  
 
The viability condition $\rho > 1$ has a gene-centric (kin-selection) interpretation in terms 
of average (neighbor modulated) fitnesses. For this purpose, 
define $\Delta p(t) = p(t+1) -p(t)$. Then it is well known that 
$\bar W \Delta p = 
p (W^A - \bar W) = 
p(1-p) (W^A - W^N)$, where $W^A$ 
and $W^N$ are the average fitnesses of types A and N, 
and $\bar W = p W^A + (1-p) W^N$ is the average fitness of all individuals.
If we choose a random type A, it will have probability proportional to $k f_k$
of being in a group with exactly $k$ types A (Bayesian sampling bias). 
Therefore $W^A = 1 + \delta \sum_{k=1}^n k f_k v^A_k / \sum_{k=1}^n k f_k$.
When $p << 1$, if we choose a random 
individual, it is likely to be in a group with no types A.
Therefore, in good approximation, 
$\bar W = 1$. Since $f(t)$ is driven towards multiples of $\nu$, 
we obtain 
\be
\Delta p \ = \ p \, \delta \, \frac{\sum_{k=1}^n k \nu_k v^A_k}
{\sum_{k=1}^n k \nu_k}, 
\label{gen}
\ee
provided $p << 1$ and $t >> 1$
(the error term is of order $p^2$ + $\beta^t$, with $0 < \beta < 1$).   
The viability condition $\rho > 1$ can also be stated as $\Delta p > 0$, in (\ref{gen}).
It is important to observe that $p(t)$ does not need to 
be monotone, and that $\Delta p > 0$ is the proper condition for invasion only 
when, as in (\ref{gen}), one is considering the stationary regime, $t >> 1$.

\section{5. Weak selection}
If selection is weak, i.e., $\delta << 1$, 
migration acts much faster than selection, providing a separation of time scales 
\cite{Rou, LKWR, RR, Lessard, Ohtsuki}. 
This allows us to replace $\nu$ in (\ref{gen}) with $\nu^0$, obtained by assuming $\delta = 0$.
Algebraic simplifications (presented in SI Appendix (Section 5))
allow us then to rewrite 
the neighbor modulated fitness relation (\ref{gen}) in the form 
\be
\Delta p \  = \ p \, \delta \, \sum_{k=1}^n \pi_k v^A_k
\label{ws}
\ee
(the error term is of order $\delta^2$),
where 
$\pi Q = \pi$ and $\sum_{k=1}^n \pi_k = 1$, \ with 
$$
Q_{i,j} =  
B_{i,j} \ + \ (1-m) \, \mbox{bin}(j-1 \, | \, n-1 \, , \, (1-m)i/n).
$$
The Markov transition matrix $Q$ and its invariant distribution $\pi$ have natural interpretations
in terms of identity by descent (IBD) under neutral genetic drift, as we explain next when we 
provide a second derivation of (\ref{ws}).

Two individuals are said to be IBD if following their lineages 
back in time, they coalesce before a migration event affects either one. 
The separation of time scales implies that when selection
acts, the demographic distribution is well approximated by that obtained in equilibrium with $\delta = 0$. 
This means that in good approximation $W^A = 1 + \delta \sum_k \pi_k v^A_k$, where $\pi_k$ is 
the $\delta=0$ equilibrium 
probability that in the group of a randomly chosen focal type A there are exactly $k$ types A
(focal included). 
But because we are supposing that types A are rare, the only individuals that are type A in this group
are those that are IBD to the focal, so that $\pi_k$ is also the probability that exactly $k$ individuals 
in this group are IBD to the focal.  
As in the derivation of (\ref{gen}), since types A are rare, we have 
$\bar{W} = 1$ and hence $\Delta p = p (W^A - 1) = \delta p \sum_{k=1}^n \pi_k v^A_k$, 
which is (\ref{ws}). 

To compute $\pi$, we will use the standard Kronecker notation $\delta_{j,i} = 1$ if $j=i$ and $\delta_{j,i} = 0$ if  $j \not= i$. 
Now, the probability $\pi_j$ that the focal is IBD to exactly $j-1$ 
other members of her group is $\delta_{j,1}$ if the focal is a migrant (probability $m$), while if she is not a migrant (probability $1-m$), then we have to consider how many individuals in her mother's group were IBD to her mother.
If, counting her mother, that number was $i$ (probability $\pi_i$, assuming demographic equilibrium)
then the probability that the focal is IBD to exactly $j-1$ other members of her group 
is equal to the probability that of the $n-1$ other members of 
her group, exactly $j-1$  are non-migrants who 
chose for mother one of the $i$ candidates 
(among $n$ possibilities) that were IBD to the focal's mother
(probability $\mbox{bin}(j-1|n-1, (1-m) i/n)$).   
Combining these pieces, we have
$$
\pi_j \ = \ m \, \delta_{j,1} \ + \ (1-m) \, \sum_{i = 1}^n \pi_i \,\mbox{bin}(j-1|n-1, (1-m) i/n). 
$$
This is exactly the same as the set of equations $\pi Q =  \pi$.

The IBD distribution $\pi$
contains all the relevant information about genetic relatedness in 
the groups, including and exceeding that given by the 
average relatedness between group members,
$R = (1-m)^2/(n - (n-1)(1-m)^2)$,
obtained from lineages, regression coefficients, or Wright's $F_{ST}$ statistics. 
Specifically, we can define $R$ as the probability that a second member chosen from the 
focal's group is IBD to the focal and then obtain (from linearity of expected values) that  
$R = ((\sum_{k=1}^n k \pi_k) - 1) / (n-1)$
is a linear function of $\pi$'s first moment. When $v^A_k$ is a non-linear function
of $k$, more information contained in $\pi$, including its  
higher moments, 
are needed to decide whether $\Delta p > 0$ in (\ref{ws}). 
It is important to also stress that 
(\ref{ws}) can be easily used for applications in 
which even the knowledge of all the moments of $\pi$ 
(see \cite{RR})
would be cumbersome to apply, 
as for instance in the computation of the short horizontal red lines in Fig.~\ref{fig2}, 
Panels A and B.  

\section{6. Large groups under weak selection}
The stationarity condition $\pi = \pi Q$ allows for a recursive computation of 
all the moments of $\pi$ 
(see SI Appendix (Section 5)).
These moments can 
then be used to 
show the powerful result that if 
$n$ is large and $m$ is small, 
then  $\pi$, when properly 
rescaled, is close to a beta distribution, with mean $R = F_{ST}$ 
(see SI Appendix (Section 6)).
In this case, if in addition to the assumption of weak selection,
also $v^A_k$ is well approximated by $\widetilde v^A_{k/n}$,
for some piecewise continuous function $\widetilde v^A_x,\ 0 \leq x \leq 1$,  then (\ref{ws}) takes the easy to apply form
\be
\Delta p \ = \ p \, \delta \, \left( \frac{1}{R} - 1 \right) 
\int_0^1 \, (1-x)^{\frac{1}{R} - 2}
\ \widetilde v^A_x \ dx, 
\label{wsln}
\ee
where $R = \frac{1}{1+2nm}$. 
Equations (\ref{gen}) and (\ref{wsln}) play complementary roles in the analysis 
of 2lFW. Both provide the condition for invasion by allele A; 
(\ref{gen}) holds in full generality, while 
(\ref{wsln}) requires special assumptions (small $\delta$, large $n$), but 
is computationally much simpler and provides a great deal of intuition, as 
we discuss next. 

Equation (\ref{wsln}) 
should be contrasted with what \cite{BR} predicted by supposing that 
the number of individuals in a group that are IBD to a focal individual would
be well approximated by a binomial with $n-1$ attempts and probability $R$ 
of success. That would lead to a normal distribution, narrowly concentrated 
close to its mean $R$, in place of the beta distribution above. 
Our result reveals a strong dependency structure among lineages, producing 
the beta distribution, with a standard deviation comparable to its mean, and a tail that 
decays slowly compared to a Gaussian distribution. 
As a consequence, fitness functions that are large only when the fraction of types A in a 
group is above a threshold value, as in the IPG, will allow for proliferation 
of types A under levels of relatedness substantially lower than that  predicted under the assumption in \cite{BR}. 
We will refer to the fact that 
the fraction of group members that are IBD to a focal individual  
has a non-vanishing standard deviation, even when selection is weak and groups are large, as {\it persistence of variability}.  
This phenomenon poses a severe limitation to the applicability of covariance-regression methods in which regression
of fitness on genotype is replaced with  derivatives, as in 
\cite{TF}, \cite{GWW} (Box 6), \cite{WGF} (condition (6.7)).  
Both the assumptions in \cite{BR}, or in \cite{TF}
applied to the IPG would have implied incorrectly that when selection is weak and 
groups are large, types A could only invade the population when rare if $R > C/B$
(these computations are presented
in 
the SI Appendix (Sections 8 and 9)). 
In a companion paper \cite{SBV} we show that
methodologies  
in which one expresses the fitness of 
a focal individual in terms of partial derivatives with respect to the focal
individual's phenotype and the phenotype of the individuals with whom the focal interacts,
as in \cite{Rou,WGF,GWW,TF, LR, LKWR}, 
require $\widetilde{v}^A_x$ to be a linear
function of $x$.

For the PG, 
(\ref{ws}) and (\ref{wsln}) clearly
reduce to the well known 
$\Delta p = p \delta (-C + BR)$.
The same is also true for the more general (\ref{gen}), as was shown in \cite{Ham75},  
where in case of strong selection the relatedness $R$ depends on the payoff functions.
In contrast, if we are under the conditions of (\ref{wsln}) with 
$\widetilde v^A_x = -C + Bx + B_2 x^2 + ... + B_l x^l$, then 
\be
\Delta p \ = \ p \, \delta \, (-C + BR + B_2 R_2 + ... + B_l R_l),
\label{wspowers}
\ee
where $R_l 
= l! R^l / [((l-1)R + 1) ((l-2)R + 1) \cdots (R+1)]$ 
is the $l$-th moment of the beta distribution. 

For the IPG,
$\widetilde v^A_x = -C + Bx$, if $0 \leq x < \alpha$, and 
$\widetilde v^A_x = (-C + Bx)T$, if $\alpha \leq x \leq 1$. 
The viability condition derived from (\ref{wsln}) can be analyzed in detail, 
by simple, but long, computations, presented in 
the SI Appendix (Section 8).
In the case $\alpha = C/B$,  
the viability condition reads $C/B - R > (T-1) R (1-C/B)^{1/R}$. When $T$ is large, 
this yields the following approximation for the critical relatedness $R = F_{ST}$:
\be
R =  
\frac{-\ln (1-C/B)}{\ln T}. 
\label{RCBlnT}
\ee
If also $C/B << 1$, then
\be
R = \frac{C/B}{\ln T} = \frac{C/B}{2.3 \, \log_{10} T}.
\label{RCBlnT+}
\ee 
The simplicity and transparency of (\ref{RCBlnT}) and (\ref{RCBlnT+}) illustrate the 
power of (\ref{wsln}), and Fig.~\ref{fig2} shows how well they compare to the more 
general, but less transparent (\ref{gen}). Note also how (\ref{RCBlnT}) and (\ref{RCBlnT+}) 
provide a direct grasp on the effect of the number of repetitions in the game, and  
a nice comparison between the PG 
and the IPG.   
Both Fig.~\ref{fig2} and (\ref{RCBlnT+}) show that 
alleles that promote contingent cooperative behavior, which 
is discontinued when participation is low, can spread under levels of genetic relatedness ( $= F_{ST}$)
more than 5 times smaller than $C/B$. 
This mechanism 
should, therefore, be seriously investigated
as a possible route for the proliferation of altruistic/cooperative behavior.  

\section{7. Conclusions}
\begin{enumerate}

\item Natural selection in group structured populations is best analyzed by a combination of group-centric 
and gene-centric perspectives and methods. Both shed light, 
carry intuition and provide computational power, in different ways.  
Rigorous mathematical analysis of models is necessary, especially when
fitness functions are non-linear, to assess the validity of non-rigorous 
approaches.

\item Contingent forms of group altruism that are discontinued when participation is low
can proliferate under biologically realistic conditions. Their 
role in the spread of altruism should be empirically investigated.

\item Natural selection can promote traits that (in net terms over a full life-cycle) 
are costly to the actors and beneficial 
to the other members of their group, under demographic conditions that are not stringent.
This can happen in large groups and with realistically high levels of gene flow. 
Excessive focus on one-shot linear public goods games in the literature has obscured this fact.

\end{enumerate}

 \begin{acknowledgments} 
The authors thank Rob Boyd for many hours of stimulating and informative conversations on the subjects in this paper. Thanks are given to Marek Biskup for assistance in exploring the nature of the distribution $\pi$. We are also grateful to Clark Barrett, Maciek Chudek, Daniel Fessler, Kevin Foster, Willem Frankenhuis, Bailey House, Laurent Lehmann, Glauco Machado, Sarah Mathew, Diogo Meyer, Cristina Moya, Peter Nonacs, Karthik Panchanathan, Susan Perry, Joan Silk, Jennifer Smith and Ming Xue for nice conversations and feedback on various aspects of this project and related subjects. This project was partially supported by CNPq, under grant 480476/2009-8.
\end{acknowledgments}

\end{article}

\end{document}